\documentclass[fleqn,twoside]{article}

\newcommand{\AmS}{{\protect\the\textfont2  A\kern-.1667em\lower.5ex\hbox{M}\kern-.125emS}}

\usepackage{espcrc2}
 
\usepackage{graphicx}
\usepackage[figuresright]{rotating}
 
\usepackage{epsfig}%,multicol,bbm}

\usepackage[normalem]{ulem}
\usepackage{graphics}
\usepackage{amsmath}
\usepackage{amsthm}
\usepackage{amssymb}
\usepackage{enumerate}
\usepackage{bm,longtable,enumerate}
\usepackage{amsmath,amssymb}

\newcommand{\Mkk}{M_{\rm KK}}

\newcommand{\ol}{\overline}

\newcommand{\AD}[1]{$\ol{\mbox{D~\,}}\!\!\!$#1}

\newcommand{\eqn}{\begin{eqnarray}}
\newcommand{\eqnx}{\end{eqnarray}}

\def\beq{\begin{equation}}
\def\eeq{\end{equation}}
\def\beqa{\begin{eqnarray}}
\def\eeqa{\end{eqnarray}}
\def\ss{\scriptscriptstyle}

\def\matt[#1,#2,#3,#4]{\left(%
\begin{array}{cc} #1 & #2 \\ #3 & #4 \end{array} \right)}

\def\v2#1{\vv2[#1]}
\def\vv2[#1,#2]{\left(%
{#1 \atop #2}\right)}

\title{Scattering vector mesons in D4/D8 model}

\author{ C. A.  Ballon Bayona\address[CBPF]
{Centro Brasileiro de Pesquisas F\'{i}sicas, \\
Rua Dr. Xavier Sigaud 150, Urca, 22290-180 Rio de Janeiro, RJ, Brazil\\}
 \thanks{email: ballon@cbpf.br},
        Henrique Boschi-Filho\address[UFRJ]{Instituto de F\'{\i}sica,
Universidade Federal do Rio de Janeiro, \\
Caixa Postal 68528, 21941-972 Rio de Janeiro, RJ, Brazil}\thanks{email: boschi@if.ufrj.br },
Nelson R.F. Braga\addressmark[UFRJ] \thanks{email: braga@if.ufrj.br} and 
Marcus A. C. Torres\addressmark[UFRJ] \thanks{email: mtorres@if.ufrj.br}}
       
\begin{document}

\begin{abstract} 
We  review in this proceedings some recent results for vector meson form factors obtained using the holographic D4-D8 brane model. The D4-D8 brane model, proposed by Sakai and Sugimoto, is a holographic dual of a semi-realistic strongly coupled large $N_c$ QCD since it breaks supersymmetry and incorporates chiral symmetry breaking. We analyze the vector meson wave functions and Regge trajectories as well.
\vspace{1pc}
\end{abstract}

\maketitle

\section{ Introduction }

Sakai and Sugimoto proposed an elegant string model dual to large $N_c$ QCD at strong coupling  \cite{Sakai:2004cn}. This model consists on the intersection of $N_c$ D4-branes and $N_f$ D8-\AD8 pair of branes in type IIA string theory in the limit $N_f\ll N_c$ where $N_c$ and $N_f$ are interpreted as the color and flavor number of strongly coupled QCD. The principal characteristic of the Sakai-Sugimoto model is the holographic description of chiral symmetry breaking $U(N_f)_L \times U(N_f)_R \to U(N_f)$ from the merging of the D$8$-\AD8 branes.  This model has been used in the recent years to describe various aspects of hadron physics \cite{Sakai:2004cn,Sakai:2005yt,Hata:2007mb,Hashimoto:2008zw,Hashimoto:2009ys}. 

In this proceedings we discuss the scattering of a photon with a vector meson using the D4-D8 brane model, summarizing the results of ref \cite{mtorres2009}. One remarkable result in the D4-D8 model is the realization of an important property of hadron physics known as vector meson dominance (VMD) \cite{Sakurai_69} where a hadron-photon  interaction is mediated by vector mesons. 
 As a consequence of VMD, the vector meson form factor takes the form of a sum involving vector meson masses and couplings.  We first present some results for the  vector meson wave functions $\psi_n(z)$, masses and couplings and then we discuss our results for the vector meson form factor. In particular, we analyze the elastic case in which we extract the magnetic and quadrupole moments. 

Form factors have also been calculated using other holographic models like the hard and soft wall model \cite{Grigoryan:2007my,Grigoryan:2007vg,Brodsky:2007hb}  and the D3/D7 brane model \cite{Hong:2003jm}.

\section{Vector mesons in the D4/D8 Model}

The induced metric in the probe D8-brane embedded in a D4 background can be written as \cite{Sakai:2004cn}
\eqn
ds^2_{\ss D8}&=& h(U_z)\, \eta_{\mu\nu} dx^\mu dx^\nu 
+ \frac49 \frac{U_{\ss{KK}}}{U_z h(U_z)} dz^2 \cr
&+& R^{3/2}U_z^{1/2} d\Omega_4^2 \, , 
\label{D8branemetricfinal}
\eqnx
\noindent where $ U_z =  U_{\ss{KK}}( 1 + z^2/U^2_{\ss KK })^{1/3}\,,\,
h(U_z) = ({U_z}/{R})^{3/2}, $ 
the constant $R$ is related to the string length $\sqrt{\alpha'}$ 
and the string coupling $g_s$ by $R^3=\pi g_s N_c \alpha'^{3/2}$ and $U_{\ss KK}$ is related to the Kaluza-Klein mass scale
by $M_{\ss{KK}} = 3 {U_{\ss{KK}}^{1/2}}/2{R^{3/2}}$. 

%The radial coordinate  $U_z$ can be rewritten in terms of $z$ by  
From the DBI action for $U(N_f)$ gauge fields in the D8-brane, we obtain a four dimensional effective lagrangian that 
can be written as \cite{Sakai:2005yt}

\beqa
{\cal L} &=&  \frac12 {\rm Tr} (\partial_\mu \tilde v_\nu^n - \partial_\nu \tilde v_\mu^n )^2 + \frac12 {\rm Tr} (\partial_\mu \tilde a_\nu^n - \partial_\nu \tilde a_\mu^n )^2 \cr
&+& M_{v^n}^2 {\rm Tr} (\tilde v_\mu^n - \frac{g_{{\cal V}v^n}}{M_{v^n}^2} {\cal V}_\mu )^2 \cr 
&+& M_{a^n}^2 {\rm Tr} (\tilde a_\mu^n - \frac{g_{{\cal A}a^n}}{M_{a^n}^2} {\cal A}_\mu )^2 \cr
&+& {\rm Tr} (i \partial_\mu \Pi + f_\pi {\cal A}_\mu )^2 + \sum_{j \ge 3} {\cal L}_j \, ,  
\label{fourdimensionalefflag}
\eeqa 
where $\tilde v_\mu^n$ and $\tilde a_\mu^n$ represent the vector and axial vector mesons, ${\cal V}_\mu$ and ${\cal A}_\mu$ are {\it external} vector and axial vector gauge fields from gauged chiral symmetry  $U(N_f)_L \times U(N_f)_R $,  $\Pi$ is a massless pion field  and ${\cal L}_j$ represent interaction terms of order $j$. Later on we turn  ${\cal A}_\mu$ off and turn on a single abelian subgroup of $U(N_f)$ in ${\cal V}_\mu$ that will be the  source of electromagnetic interaction.   
The masses and couplings are defined by
\beqa
M_{v^n}^2 = \lambda_{2n-1} M^2_{\ss{KK}} \quad ,\quad  M_{a^n}^2 = \lambda_{2n} M^2_{\ss{KK}} \cr 
g_{{\cal V}v^n} = \kappa M_{v^n}^2 \int d \tilde z \,K(\tilde z)^{-1/3} \psi_{2n-1}(\tilde z) \cr 
\!\!\!\!g_{{\cal A}a^n} = \kappa M_{a^n}^2 \int d \tilde z K(\tilde z)^{-1/3} \psi_{2n}(\tilde z) \psi_0(\tilde z)
\eeqa
where the wave functions $\psi_n$ are subjected to the conditions 
\beqa
&&\kappa \int d \tilde z (K( \tilde z))^{-1/3} \psi_n (\tilde z) \psi_m (\tilde z) = \delta_{nm} \, \, ,  \label{vectormesonnorm} \\
&&- (K(\tilde z) )^{1/3} \partial_{\tilde z} \left[ K(\tilde z) \partial_{\tilde z} \psi_n (\tilde z) \right] = \lambda_n  \psi_n (\tilde z)\, ,  \hskip 0.5cm \label{vectormesoneq}
\eeqa
where $\tilde z= z/U_{\ss KK }$, $K(\tilde z) = 1 + \tilde z^2$ and $(6 \pi)^{3} \kappa =  g_{YM}^2 N_c^2$.

Note that the constant $g_{{\cal V}v^n}$ in (\ref{fourdimensionalefflag}) is the coupling of the interaction between a  vector meson $\tilde v_\mu^n$ and an external $U(1)$ field ${\cal V}_\mu$.
\\

\begin{figure}
\centering
\includegraphics[width=5.2cm,angle=0]{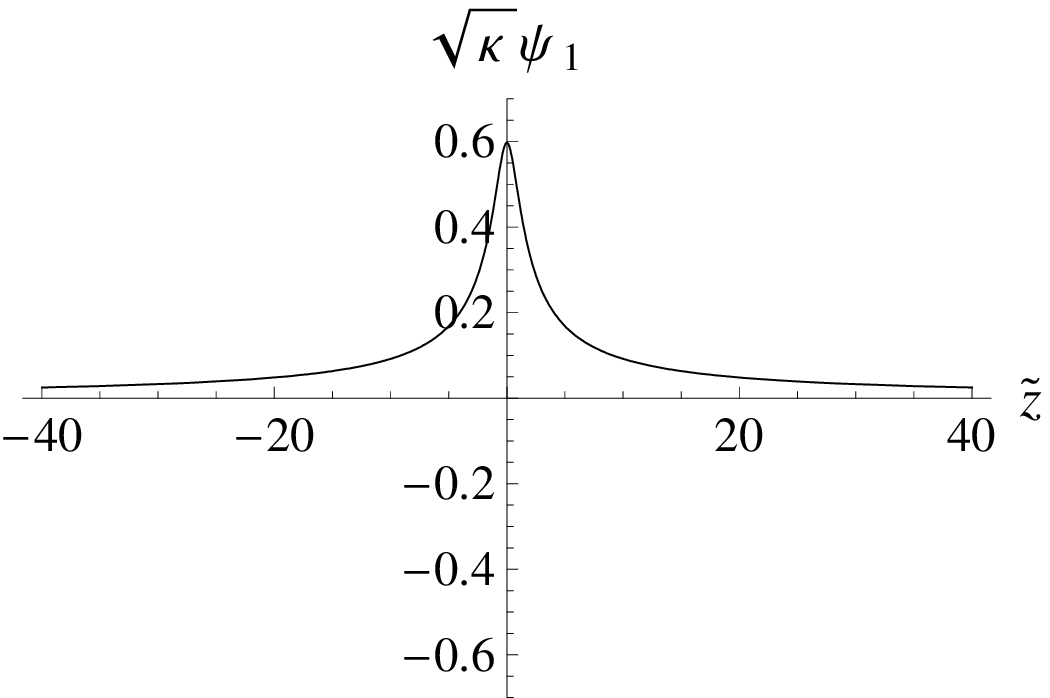}
\smallskip
\includegraphics[width=5.2cm,angle=0]{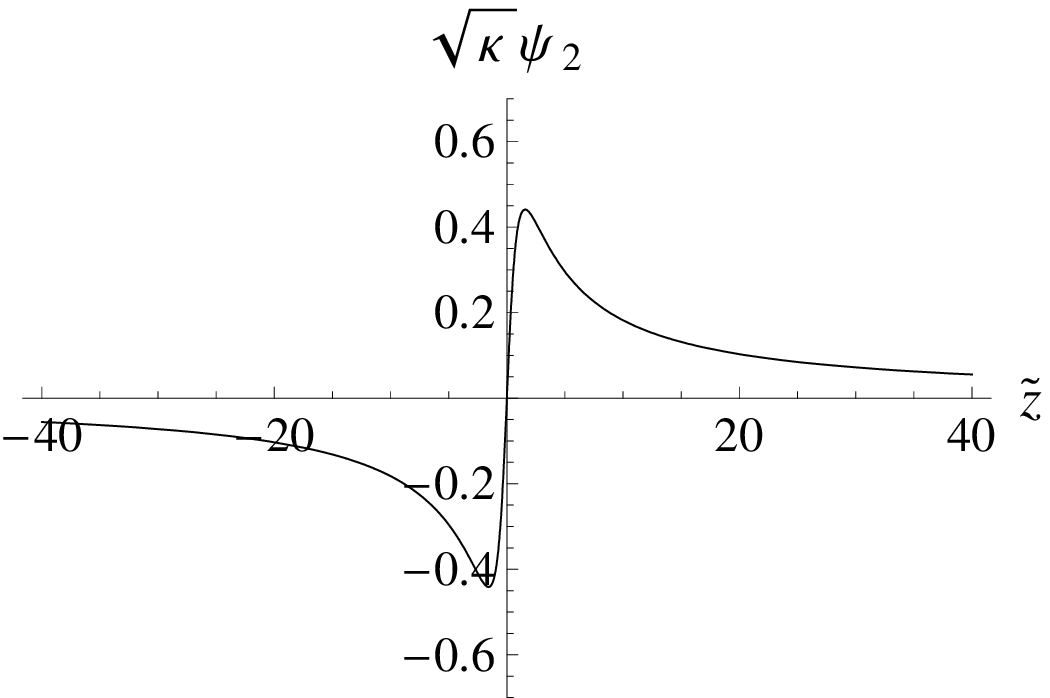}
\smallskip
\includegraphics[width=5.2cm,angle=0]{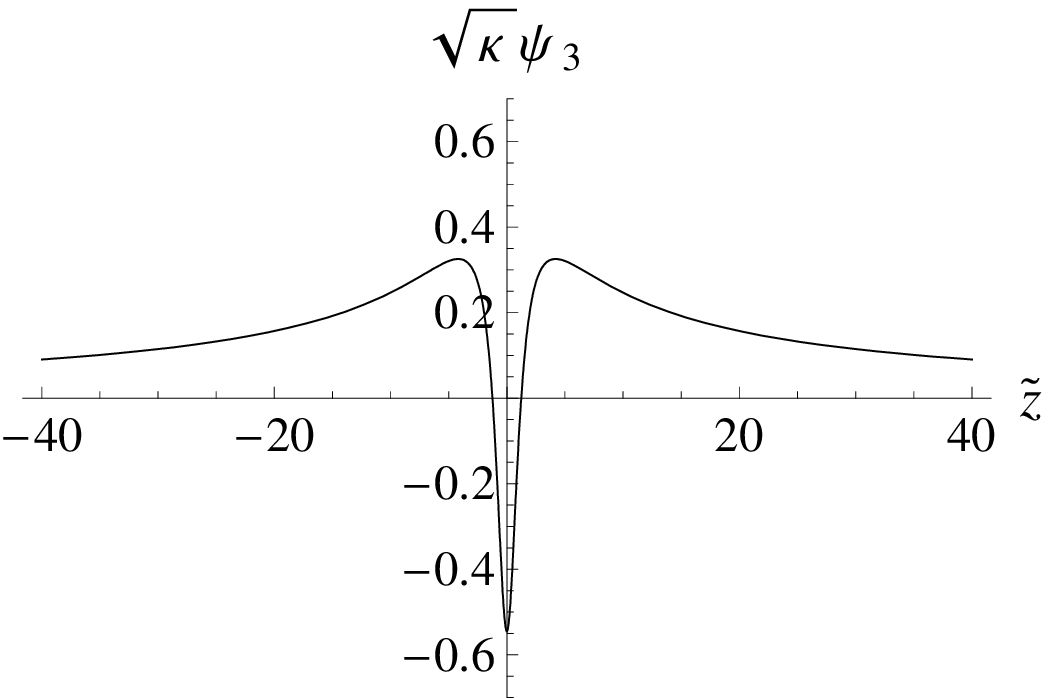}
\smallskip
\includegraphics[width=5.2cm,angle=0]{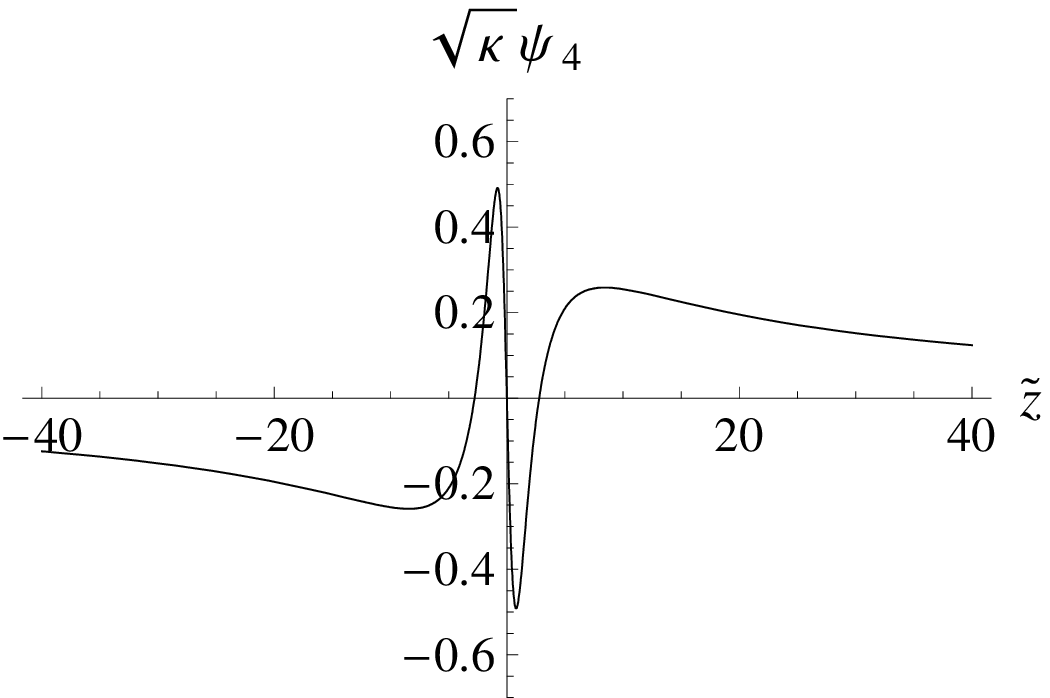}
\smallskip
\includegraphics[width=5.2cm,angle=0]{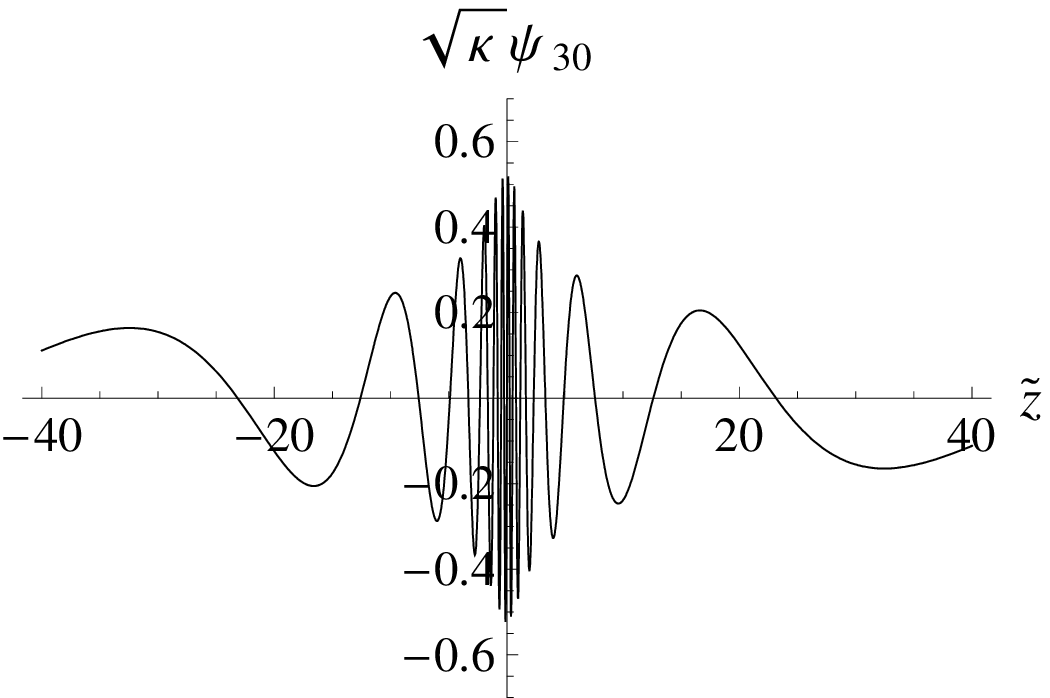}
\caption{Wave functions $\psi_{n}(\tilde z)$ multiplied by $\sqrt{\kappa}$ for the cases $n=1,2,3,4$ and $n=30$.}
 \label{wavefunctions}
\end{figure}

{\bf Wave functions.} A regularity condition for the wave functions $\psi_{2n}(\tilde z)$ and $\psi_{2n-1}(\tilde z)$ 
at the origin $\tilde z=0$, together with their parity 
\beq
\psi_{n}(-\tilde z)= (-1)^n \psi_{n}(\tilde z) \label{zparity}  
\eeq
leads to the conditions 
\beq
\partial_{\tilde z} \psi_{2n}(0) \,=\, 0 \quad , \quad \psi_{2n-1}(0)\,=\, 0 \,. 
\eeq
 
We solve numerically the equations of motion for the vector and axial-vector modes using the shooting-method. 
From the normalization condition (\ref{vectormesonnorm}) and equation of motion (\ref{vectormesoneq}) one sees 
that $\psi_n$ decrease as $\tilde z^{-1}$ when $\tilde z \to \pm \infty$. Defining 
$\tilde \psi_n \equiv \tilde z \psi_n$, the equation of motion takes the form 
\beq
\tilde z \partial_{\tilde z} \left[ \tilde z \partial_{\tilde z} \tilde \psi_n \right] \,+\, A(\tilde z) \, \tilde z \partial_{\tilde z} \tilde \psi_n \,+\, B(\tilde z) \, \tilde \psi_n \,=\, 0 \, , \label{eqvectormesontilde}
\eeq
where $A(\tilde z)$ and $B(\tilde z)$ are well behaved for large $\tilde z$:
\beqa
 A(\tilde z)=- \frac{1 + 3 \tilde z^{-2}}{1 + \tilde z^{-2}} \, 
\equiv \, \sum_{\ell=0}^{\infty} A_\ell \, \tilde z^{-2 \ell /3} \, , \hskip 1cm \cr
B(\tilde z) = 2 \frac{\tilde z^{-2}}{1 + \tilde z^{-2}} \,+\, \lambda_n \tilde z^{-2/3} (1 + \tilde z^{-2} )^{-4/3} \cr
\equiv \sum_{\ell=0}^{\infty} B_\ell \, \tilde z^{-2 \ell /3} \,. \hskip 3cm 
\label{coeffexpansions}
\eeqa
 Using the ansatz $\tilde \psi_n (\tilde z) = \sum_{\ell=0}^\infty \alpha_\ell \, \tilde z^{-2 \ell /3} $ 
one finds 
\beqa
\alpha_\ell &=& \left( \frac49 \ell^2 + \frac23 \ell  \right)^{-1}  \cr
&\times& \left [ \frac23 \sum_{k=1}^{\ell-1} k \alpha_k A_{\ell-k} 
- \sum_{k=0}^{\ell-1} \alpha_k B_{\ell-k} \right ] \hskip .9cm
\eeqa
with $\alpha_0=1$ and $\alpha_1=-(9/10) B_1$. Imposing the parity condition and the asymptotic behavior, we find numerically the wave functions $\psi_n(z)$ and their eigenvalues $\lambda_n$. If we go to the limit of small $z$, $\psi_n$ simplifies to $\sin\sqrt{\lambda_n}z$ or $\cos\sqrt{\lambda_n}z$, according to their parity. 
For large values of $\lambda_n$, there will be plenty of oscillations before the wave functions $\psi_n(z)$ reach the  $z^{-1}$ behavior. We plot some wave functions in Fig. \ref{wavefunctions}.

\begin{table}
%\\[1ex]
%Coupling constants \\[1ex]
\bigskip\centerline{\begin{array}[b]{|c|c|c|c|c|c|}
\hline
 & & & &  \\
n & \lambda_{2n-1} & \frac{g_{v^n}}{\sqrt{\kappa}M_{KK}^2} & \sqrt{\kappa}g_{v^nv^1v^1} &\lambda_{2n} \\
[0.5ex]
& & & &  \\ 
\hline\hline
1& 0.66931 &2.10936  & 0.44658&1.56877 \\
2&2.87432&9.10785 &-0.14654&4.5461 \\
3&6.59118& 20.7957& 1.8434{\ss\times}10^{\ss -2}&9.00797 \\
4&11.79669&37.1502& -3.6885{\ss\times}10^{\ss -4}&14.9573 \\
5&18.48972& 58.1701 &2.6953{\ss\times}10^{\ss -4}&22.394 \\
6&26.67017&83.834& 3.0775{\ss\times}10^{\ss -5}&31.3182 \\
7&36.33796& 114.152& 1.8572{\ss\times}10^{\ss -5}&41.7297 \\
8&47.49318&148.103& 6.9961{\ss\times}10^{\ss -6}&53.6285 \\
9& 60.1312 &188.695&3.5081{\ss\times}10^{\ss -6}&67.0149 \\ 
\hline
\end{array}}
\caption{ Dimensionless squared masses and coupling constants, where 
$\kappa= (g_{YM} N_c)^2/216 \pi^3$.}
\label{tableconstants}
\end{table}

\medskip
 
\begin{figure}
\centering
\includegraphics[width=6.7cm,angle=0]{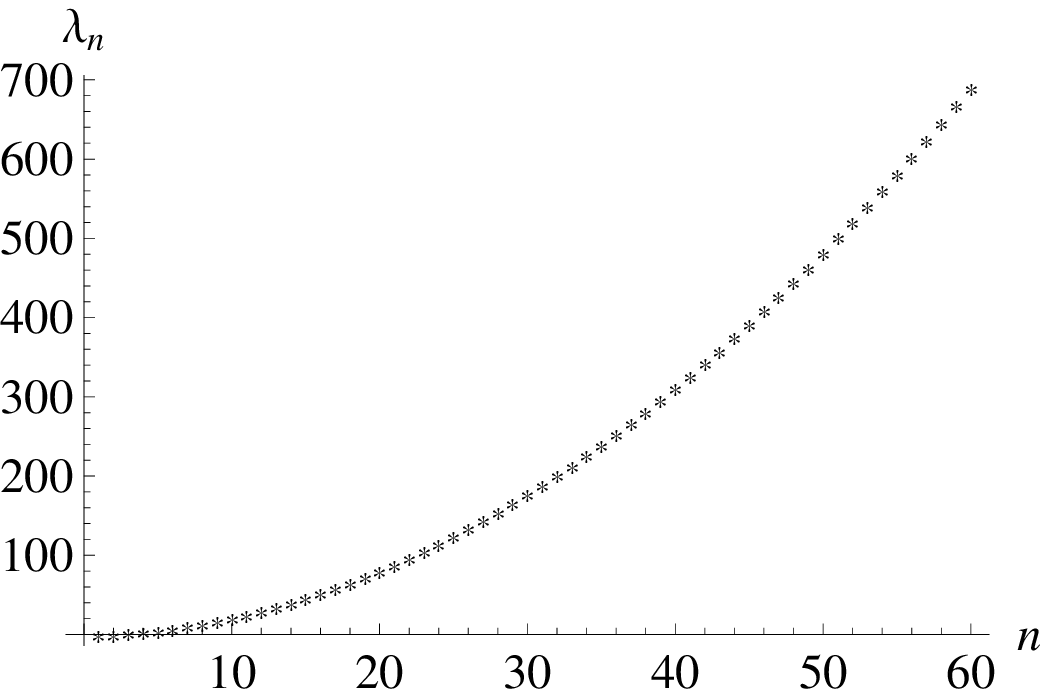}
\bigskip
\includegraphics[width=6.7cm,angle=0]{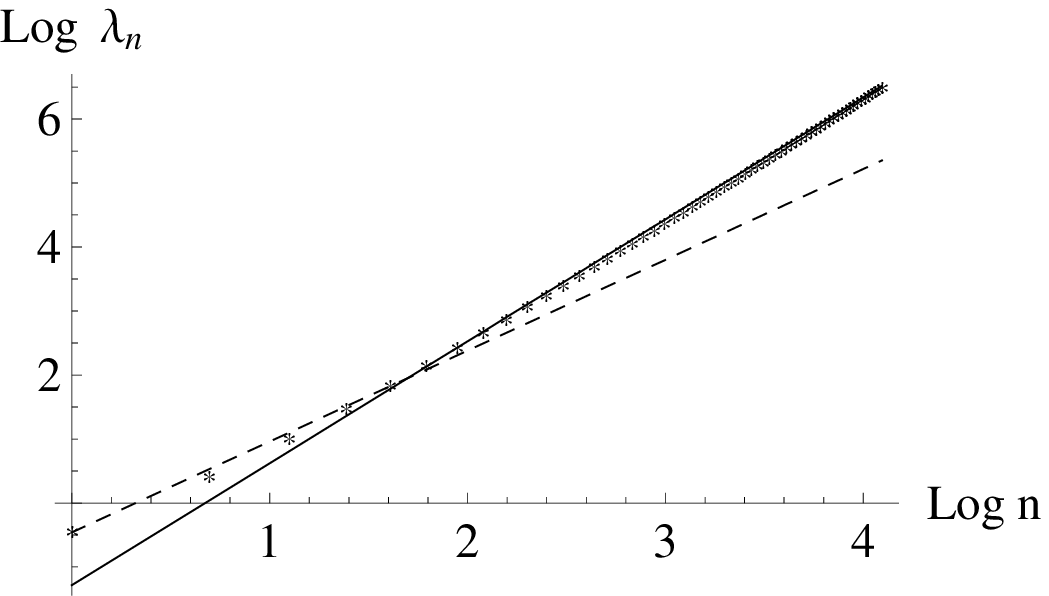}
\caption{Meson Regge trajectory for D4/D8 model. The first graph shows the dependence of $\lambda_n$ with the radial number $n$. In the logarithmic graph we fit our results, the dashed line is $ -0.46 + 1.42 \, {\rm log}\, n$ corresponding to $n=\{1,5\}$  while the solid line is  $ -1.28 + 1.91 \, {\rm log}\, n$ corresponding to $n=\{6,60\}$. }
\label{Regge}
\end{figure}

{\bf Masses and couplings.} 
Using numerical methods, we calculated eigenvalues $\lambda_n$ for $n=1$ to 60. The data for several eigenvalues 
are shown in Table \ref{tableconstants} together with the calculated coupling constants for the vector mesons. 
With such data we obtain the corresponding Regge trajectory for the vector and axial vector mesons in the D4-D8 model. The result is shown in Figure \ref{Regge}. This Regge trajectory  is different from that found for hadrons in the hard and soft wall holographic models 
\cite{deTeramond:2005su,Erlich:2005qh,BoschiFilho:2005yh,Karch:2006pv}.

%%%%%%%%%%%%%%%%%%%%%%%%%%%%%%%%%%%%%%%%%%%%%%

\section{Form Factors}
\label{Formfactors}

The Sakai-Sugimoto D4/D8 model realizes vector meson dominance (VMD) in electromagnetic scattering. In \cite{Sakai:2005yt}  they were able to show that a photon-meson-meson couplings are all cancelled, and only a dipole interaction photon-vector meson coupling survives.

Exploring further the work done in \cite{Sakai:2005yt}, we calculate the first excitation vector meson, $\rho$(770), elastic and non-elastic form factors derived from photon-meson scattering. We also calculated the axial vector mesons form factors \cite{mtorres2009}.

 \subsection{Generalized form factors}

 %%%%%%%%%%%%%%%%%%%%%%%%%%%%%%%%%%%%%%%%%% Figura %%%%%%%%%%%%%%%%%%%%%%%%%%
\begin{figure}
\begin{center}
\vskip 3.cm
\begin{picture}(30,0)(12,0)
\setlength{\unitlength}{0.06in}
\rm
\thicklines 
%%%%%%%%%%%%%%%%%%%%%%%%%%%%%%%%%%%%%%%%% Mesons Vetoriais %%%%%%%%%%%%%%%%%%%%%%
\put(-12,17){$v^m$}
\put(-9,15){\line(1,-1){7.5}}
\put(-8.6,15.4){\line(1,-1){7.5}} 
\put(-12,-3){$v^\ell$}
\put(-9,0){\line(1,1){7.5}}
\put(-8.6,-0.4){\line(1,1){7.5}}
\put(10.2,7.7){\circle*{2}}
\put(-1.2,8.1){\line(1,0){10.5}}
\put(-1.2,7.5){\line(1,0){10.5}}
\put(-1.2,7.7){\circle*{2}}
\put(4,10){$v^n$}
%%%%%%%%%%%%%%%%%%%%%%%%%%%%%%%%%%%%%%%%%%%%%%%%%  Foton  %%%%%%%%%%%%%%%%%%%%%
\put(17,10.5){$\gamma$}
\bezier{300}(10.5,7.5)(12.2,9.7)(13,7.5)
\bezier{300}(13,7.5)(14,6.5)(15,7.5)
\bezier{300}(15,7.5)(16.5,9.7)(17.5,7.5)
\bezier{300}(17.5,7.5)(18.5,6.5)(19.5,7.5)
\bezier{300}(19.5,7.5)(20.5,9.7)(21.5,7.5)
\bezier{300}(21.5,7.5)(22.5,6.5)(23.5,7.5)
\bezier{300}(23.5,7.5)(24.5,9.7)(25.5,7.5)
\end{picture}
\vskip 1.cm
\parbox{2.7 in}{\caption{Feynman diagram for vector meson form factor.}}
\end{center}\label{Feynman}
\end{figure}
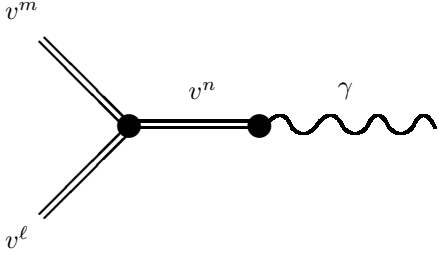
\vskip .5cm
%%%%%%%%%%%%%%%%%%%%%%%%%%%%%%%%%%%%%%%%%%%%%%%%%%%%%%%%%%%%%%%%%%%%%%%%%%%%%%%%%%

The form factors are calculated from the matrix elements of the eletromagnetic current. 
The interaction of a vector meson $v^{m}$ with momentum $p$ and polarization $\epsilon$ 
with an off-shell photon with momentum $q=p'-p$ is described by the matrix element

\begin{eqnarray}
\langle v^{m\,a}(p), \epsilon \vert {\tilde J}^{\mu c}(q) \vert v^{\ell \,b}(p'), \epsilon' \rangle 
=  \delta^4(p'-p-q) \cr\cr
\times (2\pi)^4 \langle v^{m\,a}(p), \epsilon \vert J^{\mu c}(0) \vert v^{\ell \,b}(p'), \epsilon' \rangle \hskip 1cm 
\end{eqnarray}

\noindent where ${\tilde J}^\mu$ is the Fourier transform of the electromagnetic current ${J}^\mu (x)$. 
This matrix element is calculated from the corresponding Feynman diagram shown in Figure 3. We find 
\begin{eqnarray}
\langle v^{m\,a}(p), \epsilon \vert J^{\mu c}(0) \vert v^{\ell \,b}(p'), \epsilon' \rangle 
= \epsilon^\nu {\epsilon'}^\rho f^{abc} \hskip 1cm\cr \cr
\times \left[ \eta_{\sigma\nu}(q-p)_\rho + \eta_{\nu\rho}(2p+q)_\sigma -  \eta_{\rho\sigma}(p + 2q)_\nu \right] \cr \cr
 \times 
\sum_{n=1}^\infty {g_{v^n}g_{v^mv^nv^\ell}} \left[ \frac{\eta^{\mu\sigma} + \frac{q^\mu q^\sigma}{M_{v^n}^2}}{q^2+M_{v^n}^2} \right] 
\hskip 1cm 
\end{eqnarray}

\noindent where $f^{abc}$ is the structure constant of $U(N_f)$ and $M_{v^n}$ is the mass of the vector meson ${v^n}$. 
Using the sum rule \cite{Sakai:2005yt}
\begin{equation}
\sum_{n=1}^\infty \frac{g_{v^n}g_{v^nv^mv^\ell}}{M_{v^n}^2} = \delta_{m\ell}
\label{sumrule}
\end{equation}
 
 \noindent we find 
 \begin{eqnarray}
\langle v^{m\,a}(p), \epsilon \vert J^{\mu c}(0) \vert v^{\ell \,b}(p'), \epsilon' \rangle 
= \epsilon^\nu {\epsilon'}^\rho f^{abc} \hskip 1cm \cr 
\times \big[ \eta_{\sigma\nu}(q-p)_\rho + \eta_{\nu\rho}(2p+q)_\sigma -  \eta_{\rho\sigma}(p + 2q)_\nu \big] \cr
 \!\!\! \times \left\{ \left( {\eta^{\mu\sigma} - \frac{q^\mu q^\sigma}{q^2}} \right) F_{v^mv^\ell}(q^2) 
+ \delta_{m\ell} \frac{q^\mu q^\sigma}{q^2} \right\} \hskip 0.5cm 
\label{Formlong}
 \end{eqnarray}
 
 \noindent where the generalized vector meson form factor is defined by 
\begin{equation}
F_{v^mv^\ell}(q^2)=\sum_{n=1}^\infty
\frac{g_{v^n}g_{v^nv^mv^\ell}}{q^2+M_{v^n}^2} \ .
\label{form:vn}
\end{equation} 

Taking into account the transversality of the vector meson polarizations: 
$\epsilon \cdot p =0 = \epsilon' \cdot p' $, we find 
 \begin{eqnarray}
 \langle v^{m\,a}(p), \epsilon \vert J^{\mu c}(0) \vert v^{\ell \,b}(p'), \epsilon' \rangle \hskip 2.5cm \cr\cr
= \epsilon^\nu {\epsilon'}^\rho f^{abc} 
\big[ \eta_{\nu\rho}(2p+q)_\sigma + 2(\eta_{\sigma\nu} q_\rho - \eta_{\rho\sigma}q_\nu) \big] \cr
 \times \left( {\eta^{\mu\sigma} - \frac{q^\mu q^\sigma}{q^2}} \right) F_{v^mv^\ell}(q^2)  \,.\hskip 2cm
\label{Formlong2}
 \end{eqnarray}
 
\noindent Note that the term  involving the factor $\delta_{m\ell}$ in eq. (\ref{Formlong}) 
did not contribute since in the elastic case ($m=\ell$) we have: $2p\cdot q + q^2 =0$.  

In a similiar way, for axial vector mesons we can calculate the form factors from the matrix element
$\langle a^{m\,a}(p), \epsilon \vert J^{\mu c}(0) \vert a^{\ell \,b}(p'), \epsilon' \rangle$. 
This corresponds to evaluating Feynman diagrams similar to Fig. 3, but with the external vector meson lines
replaced by the axial vector mesons $a^m$ and $a^\ell$. Note that the internal vector meson line $v^n$, 
representing vector meson dominance, is unchanged. Thus, the generalized axial vector meson form factor is
\begin{equation}
F_{a^ma^\ell}(q^2)=\sum_{n=1}^\infty
\frac{g_{v^n}g_{v^na^ma^\ell}}{q^2+M_{v^n}^2} \ .
\label{form:an}
\end{equation}

\subsection{Elastic case}

The elastic form factor for vector mesons can be obtained considering the previous calculation with the 
same vector meson $v^m$ in the initial and final states. Then, from eq. (\ref{Formlong2}) we find 

\begin{eqnarray}
 \langle v^{m\,a}(p), \epsilon \vert J^{\mu c}(0) \vert v^{m \,b}(p'), \epsilon' \rangle \hskip 2cm \cr 
 = f^{abc} \big\{ (\epsilon \cdot \epsilon') (2p+q)^\mu  \hskip 3cm \cr
 + 2\left[ \epsilon^\mu (\epsilon'\cdot q) - {\epsilon'}^\mu (\epsilon \cdot q) \right] \big\}
F_{v^m}(q^2) \,, \hskip 1cm
\label{Formelastic}
 \end{eqnarray}

\noindent where $F_{v^m}(q^2)$ is the elastic form factor:
\begin{equation}
F_{v^m}(q^2)=\sum_{n=1}^\infty
\frac{g_{v^n}g_{v^nv^mv^m}}{q^2+M_{v^n}^2} \ .
\label{form:vnelast}
\end{equation} 

\begin{figure}
\centering
\includegraphics[width=6.45cm,angle=0]{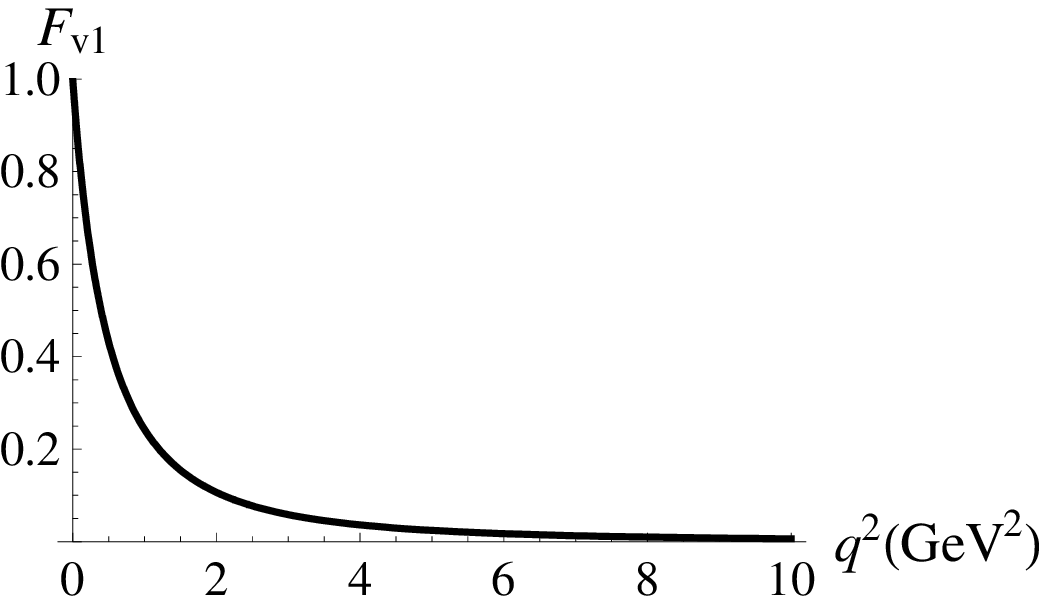} 
\includegraphics[width=6.45cm,angle=0]{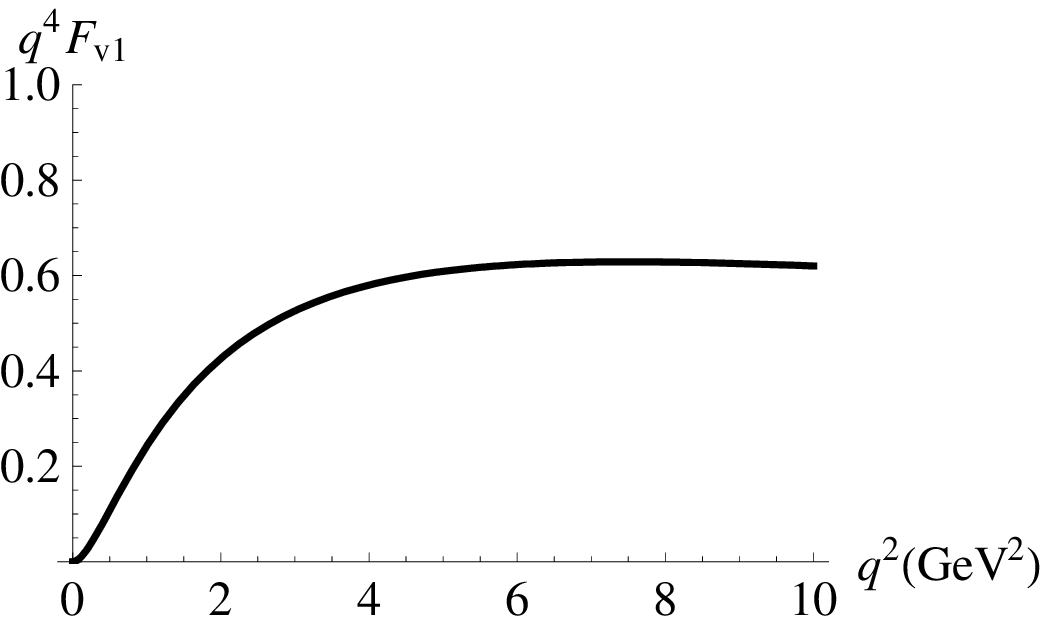}
\caption{ Elastic form factor for the $\rho$ meson.}
\label{v1formfactor}
\end{figure}

\noindent Note that eqs. (\ref{Formelastic}) and (\ref{form:vnelast}) are also valid for 
the axial vector mesons, replacing $v^m$ by $a^m$. 
Using the data from Table \ref{tableconstants}, we calculate the elastic form factors for the vector meson 
$\rho$(770) ($v^1$) and axial vector meson $a_1$(1260) ($a^1$). We plot the results in Figure 
\ref{v1formfactor}. Note that when $q^2\to 0$, the vector and axial vector form factors go to one, 
thanks to the sum rule (\ref{sumrule}).

For large values of $q^2$ we found that the form factors decrease approximately as $q^{-4}$.  
This can be seen in the second panel in Figure \ref{v1formfactor}  where we plot the form 
factors multiplied by $q^4$. 
In order to explain this behavior, we expand the elastic form factors in powers of $q^{-2}$:  
\begin{eqnarray*}
F_{v^m}(q^2) = \sum_{n=1}^\infty \frac{g_{v^n}g_{v^nv^mv^m}}{q^{2}} 
\Big( 1 - \frac{M_{v^n}^2}{q^2} + {\cal O}(\frac{1}{q^{4}}) \Big).
\end{eqnarray*} 

\noindent From this expansion we see the that the $q^{-4}$ term dominates for large $q$ if the following 
conditions hold
\begin{equation}
\sum_{n=1}^\infty
{g_{v^n}g_{v^nv^mv^m}} = 0 \,.
\label{superconv}
\end{equation}

\noindent These conditions are known as superconvergence relations. 
For the case of $v^1$ we found numerically  
\begin{equation} 
\sum_{n=1}^9
{g_{v^n}g_{v^nv^1v^1}}  \approx -0.0007889(\Mkk)^2\,.
\end{equation}

\noindent These results indicate that the conditions (\ref{superconv})  are satisfied in the D4-D8 model.

%%%%%%%%%%%%%%%%%%%%%%%%%%%%%%%%%%%%%%%%%%%%%%%%%%%%%%%%%%%%%%%%%%%%%%

\bigskip

\noindent {\bf Electric, magnetic and quadrupole form factors.} 
The matrix element of the electromagnetic current for a spin one particle in the elastic case  
can be decomposed as\cite{Grigoryan:2007vg}
\begin{eqnarray}
\langle p, \epsilon \vert J_{EM}^{\mu }(0) \vert p', \epsilon' \rangle 
=   (\epsilon \cdot \epsilon') (2p+q)^\mu F_1 ( q^2 )  \cr \cr 
+ \big[ \epsilon^\mu (\epsilon'\cdot q) - {\epsilon'}^\mu (\epsilon \cdot q) \big] 
\left[ F_1 (q^2) + F_2(q^2) \right]  \nonumber\\
+  \frac{1}{p^2} ( q\cdot \epsilon') (q \cdot \epsilon ) (2p+q)^\mu F_3 (q^2) \,.\hskip 1cm 
\label{FormDecomp}
 \end{eqnarray}
 
\noindent From $F_1 , F_2 $ and $F_3$ we can define the electric, magnetic, and quadrupole form factors:
\eqn
F_E &=& F_1 + \frac{q^2}{6 p^2} \Big[ F_2 - ( 1 - \frac{q^2}{4 p^2})  F_3 \Big] \cr\cr
F_M &=& F_1 + F_2 \cr  
F_Q &=& - F_2 + \Big( 1 - \frac{q^2}{4p^2} \Big) F_3 
\eqnx

\noindent 
Then, using eqs. (\ref{Formelastic}) and (\ref{FormDecomp}) we find that for a vector meson $v^m$
\begin{equation}
F_1^{(v^m)} = F_2^{(v^m)} =  F_{v^m}\,\,,\,\,\;\;\;\; F_3^{(v^m)} = 0 \,,\quad 
\end{equation}

\noindent where $F_{v^m}$ is given by eq. (\ref{form:vnelast}). 
Hence the electric, magnetic and quadrupole form factors predicted by the D4-D8 brane model are
\eqn
F_E^{(v^m)} = ( 1  + \frac{q^2}{6 p^2} ) F_{v^m}
\,\,\, , \,\,\,
F_M^{(v^m)} = 2 F_{v^m} \,,
\,\,\,  \,\,\, \cr
F_Q^{(v^m)} = - F_{v^m}  \,. \hskip 4cm
\eqnx

\noindent 
We can now estimate three important physical quantities associated with 
the vector mesons:  the electric radius, the magnetic and quadrupole moments.

%\noindent{\bf Electric radius}

The electric radius for the vector mesons are given by  
\begin{equation}
\langle r^2_{v^m}\rangle = -6\frac{\mathrm{d}}{\mathrm{d}q^2}F_E^{(v^m)}(q^2)|_{q^2=0} \,.
\end{equation}

\noindent Using our numerical results for the form factors of the lowest excited state $\rho$,
we find its electric radius:
\begin{equation}
 \langle r^2_{\rho}\rangle = 0.5739 \,{\rm fm}^2\,.
\end{equation}

%%%%%%%%%%%%%%%%%%%%%%%%%%%%%%%%%%%%%%%%%%%%

%\bigskip 

%\noindent{\bf Magnetic and quadrupole moments }

The magnetic and quadrupole moments are defined by
\begin{equation}
\mu  \equiv  F_M(q^2)|_{q^2=0}\,,\;\;
D  \equiv  - \frac{1}{p^2} F_Q(q^2)|_{q^2=0}.
\end{equation}

\noindent Using the fact that $F_{v^m}$  goes to one when $q^2 \to 0$, we obtain
\begin{eqnarray}
\mu_{v^m} = 2  \,,\;\;\;\;
D_{v^m} = - \frac{1}{M_{v^m}^2} .
\end{eqnarray}

Our results for electric radius, magnetic and quadrupole moments for the vector meson  $\rho$ 
are in agreement with the hard wall model results found in \cite{Grigoryan:2007vg}. 

\section{Conclusion}

We have seen that the Sakai-Sugimoto D4-D8 brane model can also be used to make predictions
on the properties of the vector mesons, in particular the $\rho$ meson. These results can be 
extended to axial vector mesons, like the $a^1$ meson \cite{mtorres2009} .

\bigskip

\noindent {\bf Acknowledgements.} We would like to thank the organizers of the LC2009 at ITA, SP, Brazil for their hospitality. The authors are partialy supported by CNPq, 
Capes and Faperj, Brazilian agencies.

\end{document}